\documentclass[aps,prd,preprintnumbers,nofootinbib,twocolumn]{revtex4}
\usepackage{bm}
\usepackage{latexsym}
\usepackage{dcolumn}
\usepackage{amsmath,amsfonts,amssymb}
\usepackage{graphicx,epsfig}
\usepackage{color}
\usepackage{amsthm}

\def\be {\begin{equation}}
\def\ee {\end{equation}}
\def\bea {\begin{eqnarray}}
\def\eea {\end{eqnarray}}
\def\bc {\begin{center}}
\def\ec {\end{center}}
\def\bfg {\begin{figure}}
\def\efg {\end{figure}}
\def\bi {\begin{itemize}}
\def\ei {\end{itemize}}

\def\beq{\begin{equation}}
\def\eeq{\end{equation}}
\def\br{\begin{eqnarray}}
\def\er{\end{eqnarray}}
\newcommand{\eel}[1] {\label{#1}\end{equation}}

\newcommand{\bdm}{\begin{displaymath}}
\newcommand{\edm}{\end{displaymath}}

\begin{document}

\title{A Proposal for Testing Gravity's Rainbow}
\author{Ahmed Farag Ali$^{1,2,3}$}\email[]{afali@fsu.edu; ahmed.ali@fsc.bu.edu.eg;}
\author{Mohammed M. Khalil$^{3}$} \email[]{moh.m.khalil@gmail.com}
\affiliation{$^1$Department of Physics, Florida State University, Tallahassee, FL 32306, USA.\\ $^2$Center for Fundamental Physics,\\Zewail City of Science and Technology, 12588, Giza, Egypt\\ $^3$Deptartment of Physics, Faculty of Science,\\Benha University, Benha, 13518, Egypt\\}
\affiliation{$^4$Department of Electrical Engineering, Alexandria University, Alexandria 12544, Egypt}

\begin{abstract}
Various approaches to quantum gravity such as string theory,
loop quantum gravity and Horava-Lifshitz gravity predict modifications of the energy-momentum dispersion relation. Magueijo and Smolin incorporated the modified dispersion
relation (MDR) with the general theory of relativity to yield a theory of gravity's rainbow.
In this paper, we investigate the Schwarzschild metric in the context
of gravity's rainbow. We investigate rainbow functions from three
known modified dispersion relations that were introduced by
Amelino-Camelia, et el. in \cite{Amelino1996pj,amelino2013,Amelino1997gz}
and by Magueijo-Smolin  in \cite{Magueijo:2001cr}. We study the effect
of the rainbow functions on the deflection of light, photon time delay,
gravitational red-shift, and the weak equivalence principle.
We compare our results with experiments to obtain upper bounds on the
parameters of the rainbow functions.
\end{abstract}

\maketitle

\section{Introduction}

The existence of the Greisen--Zatsepin--Kuzmin limit (GZK limit) as an
upper limit on the energy of cosmic rays has been regarded as a
key in the search for explorations of physics in the energy realms
that require new theories of quantum gravity which predict
events at the Planck scale. The limit is at the same order of
magnitude as the upper limit for energy at which cosmic rays have
experimentally been detected from ``distant'' sources. The limit is $5\times10^{19}$ eV \cite{Greisen:1966jv}.
In 2010, the Pierre Auger Collaboration and the High Resolution Fly's
Eye (HiRes) experiment reconfirmed earlier results of the GZK cutoff
\cite{Abraham:2010mj}. These observations predict the existence of a natural energy cutoff, and hence a modification in the dispersion relation to express the existence of the energy cutoff.

The GZK has been regarded as a key for quantum theory of gravity.
Theoretically, there are various approaches to quantum gravity that predict the existence of a minimal observable length , i.e.
the Planck length. This minimal length works as a natural cutoff, and
hence there is a departure from the relativistic dispersion relation
by redefining the physical momentum at the Planck scale.
These approaches include spacetime discreteness \cite{'tHooft:1996uc},
spontaneous symmetry breaking of Lorentz invariance in string field
theory \cite{Kostelecky:1988zi}, spacetime foam models \cite{Amelino1997gz} and
spin-network in Loop quantum gravity (LQG) \cite{Gambini:1998it}. Besides,
there are other approaches such as non-commutative geometry \cite{Carroll:2001ws}
which predicts a Lorentz invariance violation at the Planck scale. Recently, a
new approach to quantum gravity was formulated by Horava in \cite{Horava:2009uw,Horava:2009if}
that predicts a modification of the dispersion relation at the scale of quantum gravity.
All these indications suggest that Lorentz violation or deformation may be
an essential property in constructing a quantum theory of gravity.

The modification of the energy-momentum dispersion relation takes the general form
\begin{equation}
{{E}^{2}}f{{\left( E/{{E}_{p}} \right)}^{2}}-{{p}^{2}}g{{\left( E/{{E}_{p}} \right)}^{2}}={{m}^{2}}.
\end{equation}
where $E_p$ is the Planck energy, and the functions $f(E/E_p)$ and $g(E/E_p)$ satisfy
\begin{equation}
\lim\limits_{E/E_p\to0} f(E/E_p)=1,\qquad \lim\limits_{E/E_p\to0} g(E/E_p)=1.
\end{equation}
One of the interesting approaches that produce the MDR is called Doubly Special Relativity (DSR) \cite{AmelinoCamelia:2000mn}. DSR can be briefly described as an extension to special relativity that includes an invariant energy scale \cite{Magueijo:2001cr}, usually assumed to be the Planck energy, in addition to the invariance of the speed of light.

In 2004, Magueijo and Smolin proposed \cite{smolin} an extension to DSR to include curvature i.e. doubly general relativity. In this approach, the geometry of spacetime depends on the energy $E$ of the particle used to probe it. Thus, spacetime is represented by a one parameter family of metrics parametrized by the ratio $E/E_p$, forming a \emph{rainbow} of metrics, and hence the name \emph{gravity's rainbow}.

The Schwarzschild metric in the framework of gravity's rainbow is represented by the following equation
\cite{smolin}
\begin{equation}
\label{eq;schmetric}
d\tau^2=\frac{1-\frac{2GM}{r}}{f(E)^2}dt^2-\frac{1}{g(E)^2}\left( \frac{dr^2}{1-\frac{2GM}{r}}+{r^2}d\Omega^2 \right)
\end{equation}
where $G$ is Newton's constant, and $d{{\Omega }^{2}}=\sin^2\theta d{{\phi }^{2}}+d{{\theta }^{2}}$. In this paper, we use the metric in the equatorial plane $\theta = \pi/2$ and $d\theta=0$. We also use natural units, in which $c=1$, $\hbar=1$, $G=6.708\times10^{-39}\text{GeV}^{-2}$ and $E_p=1/\sqrt{G}=1.221\times10^{19}\text{GeV}$.

We investigate in this paper the implications of the modified metric \eqref{eq;schmetric}.
The choice of the Rainbow functions $f(E/E_p)$ and $g(E/E_p)$ is very important for making predictions.
That choice is preferred to be based on phenomenological motivations. Among different arbitrary choices in \cite{Garattini:2011hy,Leiva:2008fd,Li:2008gs,Ali:2014cpa,Ali:2014xqa,Awad:2013nxa,Barrow:2013gia,Liu:2007fk}, many aspects of the theory have been studied with FRW universe and black hole thermodynamics.
Many proposals exist in the literature; we will use the following three forms which
has phenomenological motivations.

\begin{enumerate}
\item The MDR proposed by Amelino-Camelia, et al. in \cite{Amelino1996pj,amelino2013}, which we will call MDR1,
\begin{equation}
f\left( E/{{E}_{p}} \right)=1,\quad g\left( E/{{E}_{p}} \right)=\sqrt{1-\eta E/E_p},
\end{equation}
which is one of the most studied MDRs, and is compatible with results from Loop Quantum Gravity and non-commutative spacetime\cite{amelino2013}.

\item The MDR proposed by Amelino-Camelia, et al. in \cite{Amelino1997gz}, which we will call MDR2,
\begin{equation}
f\left( E/{{E}_{p}} \right)=\frac{{{e}^{\alpha E/Ep}}-1}{\alpha E/{{E}_{p}}},\quad g\left( E/{{E}_{p}} \right)=1,
\end{equation}
which was proposed to explain the hard spectra from gamma-ray bursters at cosmological distances.

\item The MDR proposed by Magueijo and Smolin in \cite{Magueijo:2001cr}, which we will call MDR3,
\begin{equation}
f\left( E/{{E}_{p}} \right)=g\left( E/{{E}_{p}} \right)=\frac{1}{1-\lambda E/{{E}_{p}}},
\end{equation}
which produces a constant speed of light, and might solve the horizon problem \cite{smolin}.
\end{enumerate}

It is normally assumed that the dimensionless parameters
$\eta,~ \alpha~ \text{and}~ \lambda $ are of the order of unity, in which case
the rainbow corrections are important only when
energies (momenta) are comparable to the Planck energy (momentum),
and lengths are comparable to the Planck length.
However, if we do not impose this condition {\it a priori}, then this may signal the
existence of a new physical scale of the order of
$\eta/E_{p} \sim \alpha/E_{p} \sim \lambda/E_{p}$. Evidently, such an intermediate length
scale cannot exceed the electroweak length scale $\sim 10^{17}/E_{p}$ (as
otherwise it would have been observed) and this implies that $\eta\sim \alpha \sim \lambda \leq 10^{17}$.

In this paper, we study the effect of the modified metric \eqref{eq;schmetric} on the deflection of light,
photon time delay, gravitational red-shift, and the weak equivalence principle.
We compare our results with experiments to obtain upper bounds on the parameters of the rainbow functions.

\section{Equations of motion}
The equations of motion for particles in the modified Schwarzschild metric \eqref{eq;schmetric} were derived in \cite{Garattini:2011hy,Leiva:2008fd}. In this section, we rederive those equations in a form that will facilitate later calculations. We will start by expressing the metric as
\begin{equation}
\label{eq;metric}
d{{\tau }^{2}}=A\left( r \right)d{{t}^{2}}-B\left( r \right)d{{r}^{2}}-\frac{{{r}^{2}}}{{{g(E)}^{2}}}d{{\phi }^{2}},
\end{equation}
where
\begin{eqnarray}
& A(r)\equiv\frac{1}{f(E)^2}\left( 1-\frac{2M}{r} \right), \nonumber \\
& B(r)\equiv\frac{1}{g(E)^2}\left( 1-\frac{2M}{r} \right)^{-1}.
\end{eqnarray}

The Lagrangian associated with the metric \eqref{eq;metric} is given by \cite[P.225]{gron2007einstein}\cite{Leiva:2008fd}
\begin{eqnarray}
\mathcal{L}=&& \frac{1}{2}g_{\mu\nu}\overset{.}{x}^\mu \overset{.}{x}^\nu \nonumber\\
=&&\frac{1}{2}\left[A(r)\overset{.}{t}^2-B(r)\overset{.}{r}^2-\frac{r^2}{g^2} \overset{.}{\phi}^2\right],
\end{eqnarray}
where the dot derivative is with respect to proper time, e.g. $\overset{.}{t}=dt/d\tau$. Since the Lagrangian does not depend on the coordinates $t$ or $\phi$, their canonical momenta are constants of motion
\begin{equation}
\label{eq;ener}
\epsilon\equiv\frac{\partial \mathcal{L}}{\partial\overset{.}{t}}=A(r)\frac{dt}{d\tau},
\end{equation}
and
\begin{equation}
\label{eq;angm}
l\equiv\frac{\partial \mathcal{L}}{\partial\overset{.}{\phi}}=\frac{1}{{{g}^{2}}}{{r}^{2}}\frac{d\phi }{d\tau },
\end{equation}
where $\epsilon$ and $l$ are identified as the energy and angular momentum per unit mass.

Dividing Eq. \eqref{eq;angm} by Eq. \eqref{eq;ener}, and solving for $d\phi/dt$
\begin{equation}
\label{eq;dphidt}
\frac{d\phi }{dt}={{g}^{2}}\frac{l}{\epsilon }\frac{1}{{{r}^{2}}}A(r).
\end{equation}
Substituting Eq.\eqref{eq;ener} and Eq.\eqref{eq;angm} in the metric \eqref{eq;metric} and solving for $dr/d\tau$
\begin{equation}
\label{eq;drdtau}
{{\left( \frac{dr}{d\tau } \right)}^{2}}=\frac{\left( {{\epsilon }^{2}}-A(r)g^2l^2/r^2-A(r) \right)}{A(r)B(r)}.
\end{equation}
Dividing the square root of Eq. \eqref{eq;drdtau} by Eq.\eqref{eq;ener}
\begin{equation}
\label{eq;drdt}
\frac{dr}{dt}=\pm {{\left( 1-\frac{A(r){{g}^{2}}}{{{r}^{2}}}{{\left( \frac{l}{\epsilon } \right)}^{2}}-\frac{A(r)}{{{\epsilon }^{2}}} \right)}^{1/2}}\sqrt{\frac{A(r)}{B(r)}}.
\end{equation}
Dividing Eq.\eqref{eq;dphidt} by Eq.\eqref{eq;drdt}
\begin{equation}
\label{eq;dphidr}
\frac{d\phi }{dr}=\pm \frac{{{g}^{2}}\sqrt{A(r)B(r)}}{r^2{{\left(\left( \frac{\epsilon }{l} \right)^{2}-A(r)\frac{g^2}{r^2}-\frac{A(r)}{l^2} \right)}^{1/2}}}.
\end{equation}

For photons ($m=0$), Eqs. \eqref{eq;drdt} and \eqref{eq;dphidr} become
\begin{equation}
\label{eq;drdt0}
\frac{dr}{dt}=\pm {{\left( 1-\frac{A(r){{g}^{2}}}{{{r}^{2}}}{{\left( \frac{l}{\epsilon } \right)}^{2}} \right)}^{1/2}}\sqrt{\frac{A(r)}{B(r)}},
\end{equation}
\begin{equation}
\label{eq;dphidr0}
\frac{d\phi }{dr}=\pm \frac{{{g}^{2}}\sqrt{A(r)B(r)}}{r^2{{\left( {{\left( \epsilon/l \right)}^{2}}-A(r){{g}^{2}}/r^2 \right)}^{1/2}}}.
\end{equation}

\section{Deflection of Light}
The deflection angle of light by the sun can be calculated from Eq.\eqref{eq;dphidr0} which gives the change in $\phi$ with respect to $r$. Suppose the distance of closest approach to the sun is $r_0$. At $r_0$, $dr/d\phi=0$ because light rays change direction, leading to
\begin{equation}
\label{r0}
{{\left( \frac{\epsilon}{l} \right)}^{2}}=\frac{A(r_0)g^2}{r_0^2}.
\end{equation}
Substituting that into Eq.\eqref{eq;dphidr0} and simplifying
\begin{equation}
\label{defl}
\frac{d\phi }{dr}=\frac{g}{r}\sqrt{B(r)}{{\left( \frac{A\left( {{r}_{0}} \right)}{A\left( r \right)}\frac{r^2}{r_0^2}-1 \right)}^{-1/2}}.
\end{equation}
But since $B\left( r \right)\propto 1/{{g}^{2}}$, Eq. \eqref{defl} is the same as the standard general relativistic result \cite[P. 189]{weinberg} with no dependence on $f(E)$ or $g(E)$. Thus, gravity's rainbow has no effect on the deflection angle of light.

In hindsight, this was to be expected, since the deflection angle is calculated from $dr/d\phi$ and the modified Schwarzschild metric can be considered as the standard metric with the change $dr\to dr/g(E)$ and $d\phi\to d\phi/g(E)$. For that reason, there is no correction to the perihelion precession from gravity's rainbow either.

It is worth mentioning that the deflection of light was calculated from an energy dependent metric by Lafrance and Myers in \cite{Lafrance:1994in}, but they used a different metric from that in Eq.\eqref{eq;schmetric}, which led to an extremely small correction to the deflection angle.

\section{Photon Time Delay}
An experiment was carried out by Shapiro et al. in \cite{Shapiro} to measure the time delay of radio waves near the sun in which they bounced a radar beam off the surface of Venus and measured the round trip travel time.

Suppose a photon starts from $r=r_1$ at the earth, gets closest to the sun at $r=r_0$, and reaches Venus at $r=r_2$. We calculate the time taken by light to travel from $r_0$ to $r$ from Eq. \eqref{eq;drdt0}, after the substitution ${{\left( \epsilon /l \right)}^{2}}=A\left( {{r}_{0}} \right){{g}^{2}}/r_{0}^{2}$, as in Eq. \eqref{r0},
\begin{equation}
\label{delay}
t\left( r,{{r}_{0}} \right)=\int_{r_0}^{r}{\left( 1-\frac{A\left( r \right)}{A\left( {{r}_{0}} \right)}\frac{r_{0}^{2}}{{{r}^{2}}} \right)}^{-1/2}\sqrt{\frac{B(r)}{A(r)}}\text{d}r,
\end{equation}
and the total time of the round trip is
\begin{equation}
T=2\left( t(r_1,r_0)+t(r_2,r_0)\right).
\end{equation}

Since $A(r)\propto 1/f(E)^2$ and $B(r)\propto 1/g(E)^2$, the effect of gravity's rainbow is to multiply the standard relativistic result by $f(E)/g(E)$
\begin{equation}
\label{TTgr}
T=T_{GR}\frac{f(E)}{g(E)},
\end{equation}
where the general relativistic result is \cite[P.203]{weinberg}
\begin{eqnarray}
T_{GR}=&& 2\sqrt{r_1^2-r_0^2}+2\sqrt{r_2^2-r_0^2} \nonumber\\
&&+4GM\left(1+\ln\left(\frac{4r_1r_2}{r_0^2}\right) \right).
\end{eqnarray}
MDR3 has no effect on Eq.\eqref{TTgr} because $f(E)=g(E)$. MDR1, however, predicts that to first order in $\eta$
\begin{equation}
T=T_{GR} \left(1+\frac{E}{2E_p}\eta\right).
\end{equation}
Thus, the excess time from the result expected in flat spacetime is
\begin{eqnarray}
\Delta T=&& T-2\sqrt{r_1^2-r_0^2}-2\sqrt{r_2^2-r_0^2} \nonumber\\
\simeq&& 4GM\left(1+\ln\left(\frac{4r_1r_2}{r_0^2}\right) \right)\nonumber\\
&&+2\left(\sqrt{r_1^2-r_0^2}+\sqrt{r_2^2-r_0^2}\right)\frac{E}{2E_p}\eta .
\end{eqnarray}

The best accuracy of measuring the delay was obtained from the travel time of radio photons from earth to the Cassini spacecraft \cite{cassini} when it was at a geocentric distance of $8.43\text{AU}=6.39\times 10^{27}\text{GeV}^{-1}$, and the closest distance of the photons to the sun was $r_0=1.6R_\odot=5.64\times 10^{24}\text{GeV}^{-1}$. In natural units, the sun's mass is $M=1.116\times10^{57}\text{GeV}$. The experiment achieved an accuracy of $2.3\times 10^{-5}$; hence constraining the parameter of MDR1 by
\begin{equation}
\frac{\delta\Delta T}{\Delta T_{GR}}=\frac{\sqrt{r_1^2-r_0^2}+\sqrt{r_2^2-r_0^2}}{2GM\left(1+\ln\left(\frac{4r_1r_2}{r_0^2}\right) \right)}\frac{E}{2{{E}_{p}}}\eta<2.3\times 10^{-5}.
\end{equation}
The experiment used a multi-frequency link with the highest frequency 34,316MHz, so using $E=\omega=1.42\times10^{-13}\text{GeV}$ we get an upper bound on $\eta$ of
\begin{equation}
\eta<1.3\times 10^{20}.
\end{equation}
MDR2 has the same effect as MDR1 to first order, which sets the same bound on $\alpha$
\begin{equation}
\alpha<1.3\times 10^{20}.
\end{equation}

This bound is larger than the one set by the electroweak
scale $10^{17}$, but not incompatible with it. Moreover, with more accurate experiments in the future, this bound may be reduced
by several orders of magnitude, in which case,
it could signal a new and intermediate length
scale between the electroweak and the Planck scale.
If, instead of radio waves, the experiment used optical light,
as in lunar ranging experiments, we can get a lower bound by
about 6 orders of magnitude.

\section{Gravitational Redshift}
Suppose that light was emitted from radius $r_1$ and received at $r_2$, by how much will the light be red-shifted? In the modified metric, Eq. \eqref{eq;schmetric}, put $dr=0$ and $d\phi=0$, and solve for $dt$. Since the time measured by a remote observer is the same for the two radii, we get
\begin{equation}
dt=\frac{f\left( {{E}_{1}} \right)d{{\tau }_{1}}}{\sqrt{1-\frac{2GM}{{{r}_{1}}}}}=\frac{f\left( {{E}_{2}} \right)d{{\tau }_{2}}}{\sqrt{1-\frac{2GM}{{{r}_{2}}}}}.
\end{equation}

For light emitted from the surface of the earth, $r_1=R$, and received at height $h$, $r_2=R+h$, the relative frequency is
\begin{equation}
\frac{\omega_2}{\omega _1}=\frac{d\tau_1}{d\tau _2}=\sqrt{\frac{1-\frac{2GM}{R}}{1-\frac{2GM}{R+h}}}\frac{f\left( {{E}_{2}} \right)}{f\left( {{E}_{1}} \right)}.
\end{equation}

In MDR1, $f(E)=1$, so it predicts no difference from the standard result. Using MDR3, expanding to first order in $\lambda$, and using $E_1=\omega_1$ and $E_2=\omega_2$
\begin{equation}
\frac{\omega_2}{\omega _1}=S\left[ 1+\left(\omega_2-\omega_1\right)\frac{1}{E_p}\lambda  \right],
\end{equation}
where
\begin{equation}
S\equiv\sqrt{\frac{1-\frac{2GM}{R}}{1-\frac{2GM}{h+R}}}.
\end{equation}
Solving for $\omega_2$ to first order in $\lambda$
\begin{equation}
\omega_2=S{{\omega }_{1}}\left[ 1+\frac{{{\omega }_{1}} }{{{E}_{p}}}\left(S-1\right)\lambda  \right].
\end{equation}
Dividing by $\omega_1$ and subtracting one
\begin{equation}
\label{redshift}
\frac{\omega_2-\omega_1}{\omega_1}=(S-1)\left(1+S\frac{\omega_1}{E_p}\lambda\right).
\end{equation}

The new term in Eq.\eqref{redshift} is proportional to the frequency. Thus, we get a greater correction using high energy photons. This is the case in the Pound-Snider experiment \cite{Pound:1965zz}, which used gamma rays of energy $14.4\times10^{-6}\text{GeV}$. The experiment was performed in a $22.86\text{m}=1.16\times 10^{17}\text{GeV}^{-1}$ tower, and achieved an accuracy of 1\%, which means that
\begin{equation}
S\frac{{{\omega }_{1}} }{{{E}_{p}}}\lambda < 0.01
\end{equation}
Using the mass and radius of the earth, $M=5.972\times 10^{24}\text{kg}=3.350\times10^{51}\text{GeV}$, and $R=6.378 \times 10^6 \text{m}=3.232\times10^{22}\text{GeV}^{-1}$, we get an upper bound on $\lambda$ of
\begin{equation}
\lambda < 8.5\times 10^{21},
\end{equation}
and on $\alpha$ of
\begin{equation}
\alpha< 1.7\times 10^{22}.
\end{equation}

These bounds are larger than the one set by the electroweak
scale $10^{17}$, but compatible with it (as was
the case for photon time delay). Repeating the experiment with current technology, and using higher energy gamma rays can lower these bounds, and perhaps predict an
intermediate length scale.

\section{The Weak Equivalence Principle}
Tests of the weak equivalence principle include free fall experiments and torsion balance experiments, which are designed to detect any dependence of gravitational acceleration on mass or composition.

To find the gravitational acceleration in gravity's rainbow, we will follow the derivation in \cite[P.3-32]{taylor2000exploring}. For a mass falling radially from rest at $r_0$, $d\tau^2=\left(1-2GM/r_0\right)dt^2/f^2$. Thus, the energy, Eq.\eqref{eq;ener}, is
\begin{equation}
\epsilon=\frac{1}{f^2}\left(1-\frac{2GM}{r}\right)\frac{dt}{d\tau}=\frac{1}{f}\sqrt{1-\frac{2GM}{r_0}}.
\end{equation}
Substituting $d\tau$ from the previous equation in the metric \eqref{eq;schmetric} with $d\phi=0$, and solving for $dr/dt$
\begin{equation}
\frac{dr}{dt}=\frac{g}{f}\left( 1-\frac{2GM}{r} \right)\sqrt{\frac{\frac{2GM}{r}-\frac{2GM}{r0}}{1-\frac{2GM}{r0}}}.
\end{equation}
For a static observer on a spherical shell of radius $r$,
\begin{equation}
dt_{sh}=\frac{\sqrt{1-2GM/r}}{f}dt, \quad dr_{sh}=\frac{dr}{g\sqrt{1-2GM/r}}
\end{equation}
Thus,
\begin{equation}
\frac{dr_{sh}}{dt_{sh}}=\sqrt{\frac{\frac{2GM}{r}-\frac{2GM}{r0}}{1-\frac{2GM}{r0}}}.
\end{equation}
Differentiating with respect to $t_{sh}$ and substituting $r=r_0$ we find that the gravitational acceleration experienced by a radially falling mass from $r_0$ is
\begin{equation}
a=\frac{{{d}^{2}}{{r}_{sh}}}{d{{t}_{sh}}^{2}}=\frac{GM}{r_0^2}{{\left( 1-\frac{2GM}{{{r}_{0}}} \right)}^{-1/2}}g(E),
\end{equation}
which is energy dependent, and reduces to the standard relativistic result when $g(E)\to1$.

For two test bodies with different energies
\begin{equation}
\frac{\Delta a}{a_1}=\frac{g(E_2)-g(E_1)}{g(E_1)}
\end{equation}
MDR2 has no effect because $g(E)=1$, but MDR1 and MDR3 to first order lead to
\begin{equation}
\label{a1}
\frac{\Delta a}{a_1}=\frac{\left| {{E}_{1}}-{{E}_{2}} \right|}{2{{E}_{p}}}\eta ,
\end{equation}
and
\begin{equation}
\label{a2}
\frac{\Delta a}{a_1}=\frac{\left| {{E}_{1}}-{{E}_{2}} \right|}{{{E}_{p}}}\lambda.
\end{equation}

Ref.\cite{Schlamminger:2007ht} reports the measurement of the acceleration of Beryllium and Titanium test bodies using a rotating torsion balance, and achieved an accuracy of $\Delta a/a=1.8\times10^{-13}$. The masses of the Beryllium and Titanium atoms are 9.0122u, and 47.867u respectively, where an atomic unit $1\text{u}=0.931\text{GeV}$. From Eqs. \eqref{a1} and \eqref{a2} we obtain the bounds
\begin{equation}
\label{bounds}
\eta<1.2\times 10^5, \qquad  \lambda<6.1\times 10^4.
\end{equation}

These bounds can be lowered further in future experiments such as the SR-POEM project \cite{Reasenberg:2012fk} in 2016, which should reach an accuracy of $\Delta a/a=2\times10^{-17}$. Thus, it might actually detect the effect of rainbow functions, or constrain their parameters to the order unity.

It should be stressed that these bounds are much more stringent than those derived in previous examples. Therefore, it could signal a new and intermediate length scale between the electroweak and the Planck scale.
Besides, we have found that even if the rainbow parameters $\sim 1$, we still might measure gravity's rainbow corrections in the weak equivalence principle case.
This is an improvement over the general conclusions of \cite{Lafrance:1994in,Garattini:2011hy}, where it was shown
that gravity's rainbow corrections are virtually negligible which make our
result new and interesting.

\begin{table*}[ht]
\caption{Bounds on the parameters of rainbow functions}
\label{Table1}
\begin{ruledtabular}
\begin{tabular}{lccc}
Experiment & $\eta$ & $\alpha$ & $\lambda$ \\
\hline Deflection of Light and Perihelion Shift & - & - & - \\  Photon Time Delay: Cassini Spacecraft & $1.3\times 10^{20}$ & $1.3\times 10^{20}$ & - \\
Gravitational Redshift: The Pound-Snider Exp. & - & $1.7\times 10^{22}$ & $8.5\times 10^{21}$ \\
Gravitational Acceleration of Br and Ti & $1.2\times 10^{5}$ & - & $6.1 \times 10^{4}$ \\
\end{tabular}
\end{ruledtabular}
\end{table*}

\section{Conclusions}

In this paper, we studied the implications of the modified Schwarzschild metric in gravity's rainbow. We derived the equations of motion, and found that gravity's rainbow predicts no corrections to the deflection angle of light or to the perihelion precession.
Besides, we have investigated the consequences of gravity's rainbow corrections
on various gravitational phenomena such as the time delay of light, gravitational redshift, and the weak equivalence principle. We have found that the upper bounds on the rainbow parameters are around $10^{20}$, $10^{22}$, and $10^{4}$ respectively. These bounds are summarized in Table \ref{Table1}.
The first and second bounds give a length scale larger than the one set by the electroweak scale $10^{17}$, but compatible with it. Moreover, with more
accurate experiments in the future, these bounds may be reduced by several orders of magnitude.
The last bound was found to be more stringent, and might be consistent with that set by the electroweak scale.
Therefore, it could signal a new and intermediate length
scale between the electroweak and the Planck scale.
On the other hand, we have found that even if the rainbow parameters $\sim 1$, we still might measure gravity's rainbow corrections in case of the weak equivalence principle.
This is in fact an improvement over the general conclusion of \cite{Lafrance:1994in,Garattini:2011hy}, where it was shown that gravity's rainbow effects are virtually negligible,
which appears to be a new and interesting result. This definitely opens a phenomenological
window for investigating the gravity's rainbow corrections with low energy systems which we hope to report
in the future.

\section*{Acknowledgments}
The research of AFA is supported by Benha University (www.bu.edu.eg) and CFP in Zewail City.

\bibliography{rainbowtests}

\end{document}